\documentclass[12pt]{article}

\usepackage[margin=1.1in]{geometry}
\usepackage{amsmath}
\usepackage{graphicx}
\usepackage{amsfonts}
\usepackage[round]{natbib}
\usepackage[colorlinks,bookmarksopen,bookmarksnumbered,citecolor=red,urlcolor=red]{hyperref}
\usepackage{tabularx,ragged2e,booktabs,caption}
\usepackage{authblk}
\usepackage{subfig}
\graphicspath{}
\usepackage{setspace}

\DeclareMathOperator*{\argmax}{arg\,max}

\newcommand{\thetahat}{\hat{\theta}}

\newcommand{\all}{\text{all}}
\newcommand{\Wald}{\text{Wald}}
\newcommand{\lssb}{\text{LSSB}}
\newcommand{\lssbw}{\text{LSSBw}}
\newcommand{\uminp}{\text{UminP}}
\newcommand{\deltahat}{\hat{\delta}}

\newcommand{\fscl}{\text{fscl}}

\newcommand{\sumkK}{\sum_{k=1}^{\Ncl}}

\newcommand{\Ncl}{N_{cl}}
\newcommand{\A}{\mathcal{A}}

\begin{document}
\doublespacing

\title{Parsimonious and powerful composite likelihood testing for group difference and genotype-phenotype association}

\author[1]{Zhendong Huang}
\author[1]{Davide Ferrari \thanks{Corresponding author: Davide Ferrari, School of Mathematics and Statistics, The University of
Melbourne, Parkville, VIC 3010, Australia. E-mail: \url{dferrari@unimelb.edu.au}.}}
\author[1]{Guoqi Qian}
\affil[1]{School of Mathematics and Statistics, The University of Melbourne}
\date{}
\maketitle

\begin{abstract}
Testing the association between a phenotype and many genetic variants from case-control data is essential in genome-wide association study (GWAS). This is a
challenging task as many such variants are correlated or non-informative. Similarities exist in testing the population difference between two groups of
high dimensional data with intractable full likelihood function. Testing may be tackled by a maximum composite
likelihood (MCL) not entailing the full likelihood, but current MCL tests are subject to power loss for involving non-informative or redundant
sub-likelihoods.  In this paper, we develop a forward search and test method for simultaneous powerful group difference testing and
informative sub-likelihoods composition. Our method constructs a sequence of Wald-type test statistics by including
only informative sub-likelihoods progressively so as to improve the test power under local sparsity alternatives.
Numerical studies show that it achieves considerable improvement over
the available tests as the modeling complexity grows.
Our method is further validated  by testing the motivating GWAS data on breast cancer with interesting results obtained.

\end{abstract}

\section{Introduction}

Testing population difference between two groups of multivariate data is common in many fields of statistical
research. Due to significant development of data acquisition technologies in recent years, more and more complex data
--- e.g. involving temporal or spatial dependence among the sample units --- can now be readily collected for statistical
analysis. However, this entails the use of tractable statistical models which are not easily available. In particular, it may be
difficult or even impossible to specify the full likelihood function for testing the group difference.
These challenges are common in analyzing case-control data in genome-wide association study (GWAS), where
for example we test associations between a binary breast cancer phenotype and
various genotype variants known as the single nucleotide polymorphisms (SNPs).
Note that testing genotype-phenotype association from case-control data can be formulated as a two-sample
statistical test problem.
But association testing for many genotype variants altogether entails a high-dimensional statistical model, and makes
it difficult to formulate a computationally tractable full likelihood  \citep{han2012composite}.

These issues naturally suggests approximating the full likelihood function by a computationally tractable one for
constructing the test statistics for association testing. A well-developed approximation is based on the maximum
composite likelihood estimator (MCLE),  obtained by maximizing the product of low-dimensional sub-likelihood objects
instead of the full likelihood.   \cite{besag1974spatial} proposed
composite likelihood estimation for spatial data while \cite{lindsay1988composite} developed composite
likelihood estimation in its generality.  Over the years, composite likelihood methods have proved useful in
many applied fields, including  geo-statistics, spatial extremes and  statistical genetics.
See \cite{varin2011overview} for a comprehensive survey on methods and applications.

Like the familiar maximum likelihood estimator (MLE), the MCLE is asymptotically unbiased and normally distributed
under regularity conditions. This feature, is beneficial for constructing Wald-type statistics for testing group differences
(see \cite{geys1999pseudolikelihood} and \cite{molenberghs2005models} among others), can also be used in
MCLE based testing. The standard approach here is to form a statistic using all the available
data-subsets (so that the MCLE is computed by combining all the feasible sub-likelihood components).
Although the resulting Wald test has known null distribution in the limit due to the
asymptotic normality of MCLE, it may exhibit unsatisfactory power when the number of parameters in the model is
moderate or large relative to the sample size.

In our view, forming a test statistic by all the available sub-likelihoods is
not always well-justified from either statistical or computational perspective. Specifically, when the noise in the data
is evident and the statistical model considered is very complex,  inclusion of sub-likelihoods that do not explain group
differences will mainly be adding noise to the Wald statistic. Clearly, this unwanted noise has the undesirable effect
of deteriorating the overall test power. A better strategy would be to choose only informative sub-likelihoods
relevant to group differences, while dropping noisy or redundant components as much as possible.

Prompted by the above discussion, we propose a new approach --- referred to as the forward step-up
composite likelihood (FS-CL) testing --- for group difference testing. Given a set of candidate data subsets used for
constructing the sub-likelihood objects, our FS-CL method carries out simultaneous testing and data noise reduction by
selecting a best set of sub-likelihoods so as to improve the resulting test power. Differently from the existing approaches,
we impose a sparsity requirement on our alternative hypothesis reflecting the notion that only certain portion of
data subsets fundamentally explains the difference between groups. While testing the null hypothesis of no difference
between groups, our method makes efficient use of data by dropping noisy or redundant data subsets to
the maximum extent. This procedure is implemented by a forward search algorithm which, similar to the well-established
methods in variable selection, progressively includes one more sub-likelihood at each step until no significant
improvement in terms of power is observed.

The new approach proposed can be extended to general linear hypothesis testing (cf. chapter 7 of
\cite{Lehmann2005testing}) without fundamental difficulty, but will not be pursued in detail in this paper.
The remainder of the paper is organized as follows. In Section~\ref{sec:framework}, we describe the main
framework for composite likelihood estimation and overview the existing Wald-type association tests.  In
Section~\ref{sec:method}, we describe the new FS-CL methodology and propose the forward search algorithm.
In Section~\ref{sec:numerical}, we study the finite-sample properties of our method in terms of Type I error probability
and power using simulated data. In Section~\ref{sec:real}, we apply our test to the case-control GWAS
data from Australian Breast Cancer Family Study. In Section~\ref{sec:conclusion}, we conclude the paper
by providing some final remarks.

\section{Composite likelihood inference} \label{sec:framework}

\subsection{Sparse composite likelihood estimation}\label{sec2.1}

Consider a random sample of $n$ observations on a $d$-dimensional random vector $Y=(Y_1, \dots, Y_d)^T$ following a probability
density function $f(y; \theta)$, with unknown parameter $\theta \in \Theta \subseteq \mathbb{R}^q$ and
$q=\dim(\Theta) \geq 1$.
Let $\thetahat(w)$ be the profiled maximum composite
likelihood estimator (MCLE) of $\theta$, obtained by  maximizing the composite likelihood function
\begin{equation}  \label{comp_lik}
\ell_{cl}(\theta; w) = \left(\sumkK w_k\right)^{-1}\sumkK  w_{k} \ell_{k}(\theta),
\end{equation}
where $\Ncl$ is the total number of sub-likelihood objects considered,
$w=(w_1, \dots, w_{\Ncl})^T \in \Omega =\{0,1\}^{\Ncl}$ is a vector of
binary weights referred to as composition rule, and
$\ell_{k}(\theta) \propto \log f(S_k;\theta)$
is the sub-likelihood defined on the $k$th data subset $S_k$. The composite likelihood
design is typically user-specified \citep{varin2011overview, Lindsay2011issues}. For example, $\ell_k$ can be  based
on all marginal events ($S_k=\{y_k \}$, $k=1,\dots,d$), all pair-wise events
($S_k=\{y_j, y_l\}$,  $1 \leq j<l\leq d$ ), or conditional events ($S_k=\{y_k|y_j, j\neq k\}$, $k=1,\dots, d$).

In our parsimonious composition framework, each sub-likelihood $\ell_{k}(\theta)$
is allowed to be selected or not, depending on whether
$w_k$ takes value $1$ or $0$, which results in an efficient use of the data. The total number of selected
sub-likelihoods, $\Vert w \Vert =\sumkK w_k$, can be much smaller than the total $\Ncl$ ones available.
This is in contrast with the frequently used composite likelihood setting where all the $\Ncl$
sub-likelihoods are selected. Particularly, in the latter case $w = w_{all} = (1,\dots, 1)^T$, and no data
noise reduction is attained.

A complication related to notations in composite likelihood is that the parameter $\theta$ does not always have
all its elements involved in each sub-likelihood $\ell_{k}(\theta)$. To facilitate presentation in the sequel, we rewrite
$\ell_{k}(\theta)$ as $\ell_k(\theta_k)$ by using $\theta_k$ to represent the parameter involved in $\ell_k(\cdot)$.
Thus the parameter $\theta$ is equivalently represented by $(\theta_1,\dots, \theta_{\Ncl})$ in composite
likelihood. This necessarily means $(\theta_1,\dots, \theta_{\Ncl})$ may contain common elements or elements
of known values. For example, if $Y$ follows a $d$-variate normal distribution $Y \sim N_d(\mu, \sigma^2 I)$
with $\mu = (\mu_1, \dots, \mu_{d-1},0)^T$ and $I$ being the identity matrix, one may define sub-likelihoods
using marginal normal distributions
$N_1(\mu_k, \sigma^2_k)$, $k=1,\dots, d =\Ncl$ and equate $\mu_d$ with 0 and all $\sigma^2_k$'s with $\sigma^2$.
In applying parsimonious likelihood composition a subset of $(\theta_1,\dots, \theta_{\Ncl})$ indexed by
the composition rule $w$ may be adequate for representing $\theta$; such a subset is denoted as $\theta(w)$ from
the subspace $\Theta_w$. It is easy to see that $\Theta_w \subseteq \Theta$ and $\dim(\Theta_w)\leq \dim(\Theta)$
although the cardinality $q_w = |\Theta_w|$ may be greater than $q=|\Theta|=\dim(\Theta)$. In the above
example of $Y \sim N_d(\mu, \sigma^2 I)$, $q=d$ and $q_{w}=2(d-1)$ if $w=(1,\dots,1,0)$.
There also exist examples where $\dim(\Theta_w)<\dim(\Theta)$.
The parameter design discussed here is often used to
simplify formulation and computation in complex models \citep{varin2011overview}.
With this in mind we regard $\theta(w)$ as representing $\theta$ or one of its sub-vectors in this paper,
and denote the effective dimension of $\theta(w)$ as $d_w$, $d_w\leq q$.

For fixed $w$, the MCLE $\hat{\theta}(w)$ based on data of sample size $n$ is a $\sqrt{n}$-consistent and asymptotically
normally distributed estimator of $\theta(w)$ under appropriate regularity conditions \citep{varin2011overview}. Specifically,
$\sqrt{n} (\hat{\theta}(w)-\theta(w))$ follows asymptotically a $d_w$-variate normal distribution
with zero mean and $d_w \times d_w$ covariance matrix $V(\theta,w)= G^{-1}(\theta,w)$, where
$G(\theta,w)=n^{-1}H(\theta,w){J(\theta,w)}^{-1}H(\theta,w)$, with $H(\theta,w)= -E[\nabla^2 \ell_{cl}(\theta; w)]$ and
$ J(\theta,w)= Var[\nabla \ell_{cl}(\theta; w)]$ being the Godambe information matrix
\citep{godambe1960optimum}.  Next, we exploit MCLE's asymptotic normality to derive sensible test
statistics for group difference testing.

\subsection{Wald-type tests for group differences} \label{sec:tests}
Let $Y_i^{(g)} \sim f(y; \theta_g)$, $i=1,\dots,n_g$ ,  $\theta_g \in \Theta\subseteq \mathbb{R}^q$,
be $d$-vector observations in two groups indexed by $g=0,1$ (e.g. case and control groups).
As just discussed in section~\ref{sec2.1}, we represent $\theta_g$ by $\theta_g(w_\all)=(\theta_{g1},\dots,\theta_{g\Ncl})$,
with each $\theta_{gj}$, $j=1,\dots,\Ncl$, being a $p$-dimensional parameter vector corresponding to the $j$th
sub-likelihood. Note that the effective dimension of $\theta_g(w_\all)$ here still equals $q$ thus some $\theta_{gj}$'s
given $g$ must contain common elements or some elements of known values.
Suppose $\theta_g(w_\all)$'s are to be estimated by MCLE. A Wald-type statistic can be naturally constructed to
test $H_{0}: \delta \equiv \theta_1 - \theta_0 = 0$ vs. $H_1:\delta\neq 0$, which is
\begin{align} \label{Wald}
T_\Wald\equiv n \ \hat{\delta}^{T} \widehat{V}^{-1} \hat{\delta} ,
\end{align}
where $\hat{\delta}  =  \hat{\theta}_{1}(w_\all)- \hat{\theta}_{0}(w_\all)$ with $w_\all =(1, \dots, 1)^T$ and
$\hat{\theta}_{g}(w_\all)$, $g=0,1$, being the MCLEs for the two groups;
and $\widehat{V} = \widehat{V}(w_\all)$  is a consistent
estimator of the asymptotic covariance matrix of $\sqrt{n}\hat{\delta}$.
It is easy to see that $\hat{\delta}$ can be regarded as an MCLE for the parameter difference of the two
groups when no sub-likelihood selection is taken.

Under the null hypothesis $H_0: \delta =   0$, the statistic $T_{\Wald}$
follows asymptotically a chi-square distribution with $q$ degrees of freedom \citep{molenberghs2005models}.
Although $T_{\Wald}$ has a known null distribution, the power  of
the test can be unsatisfactory when $q$ is relatively large. This is due to the fact that with no selection of
sub-likelihood components, pronounced noise in data subsets that does not explain the difference  between groups
may deteriorate Wald test's power as a consequence of inflating the covariance matrix $\widehat{V}$.

To mitigate the above issues, \cite{han2012composite} studied modifications of the basic Wald test.
They replaced $\widehat{V}$ by simpler matrices resulting in two test statistics they called
$T_{\lssb}\equiv n \hat{\delta}^{T} \hat{\delta}$
and $T_{\lssbw}\equiv n\hat{\delta}^{T} \text{diag}(\hat{V})^{-1} \hat{\delta}$,
 where $\text{diag}(\hat{V})$ denotes the diagonal matrix of $\hat{V}$. The
asymptotic null distributions for both statistics have the form
$\sum_{j=1}^{q}\tau_{j}X_{j}^{2}$, where $X_{1}^2,\dots,X_q^2$ are independent chi-square random variables
with $1$ degree of freedom, and $\tau_{j}$ denotes the $j$th eigenvalue of
$\hat{V}$ and $\hat{V} \text{diag}(\hat{V})^{-1}$, for the LSSB test and LSSBw test,
respectively. Following \cite{zhang2005approximate}, the distribution of
$\sum_{j=1}^{q}\tau_{j}X_{j}^{2}$ can be approximated by a scaled-shifted
chi-square distribution for $a\chi_{r}^{2}+b$, where $\chi_{r}^{2}$ is random variable having a chi-square
distribution with $r$ degrees of freedom, with $a$, $b$ and $r$ given by
\begin{align*}
a=\frac{\sum_{j=1}^q\tau_{j}^{3}}{\sum_{j=1}^q\tau_{j}^{2}},\;\;\;\;\;
b=\sum_{j=1}^q\tau_{j}-\frac{(\sum_{j=1}^q\tau_{j}^{2})^{2}}{\sum_{j=1}^q\tau_{j}^{3}},
\;\;\;\;\;r=\frac{(\sum_{j=1}^q\tau_{j}^{2})^{3}}{(\sum_{j=1}^q\tau_{j}^{3})^{2}}.
\end{align*}
The LSSB and LSSBw statistics are easier to compute compared to (\ref{Wald}). But they may still have
low power when the sample size is not large enough. Another common test  is the UminP test with test statistic
$T_{\uminp}= \max_{\substack{1\leq j\leq q}}\{\sqrt{n}|\hat{\delta}|_{j}/{\hat{V}^{1/2}_{jj}} \}$
where $\hat{V}_{jj}$ is the $j$th diagonal element of $\hat{V}$ \citep{pan2009asymptotic}.  Other test statistics
have been derived from the composite likelihood ratio (CLR) test and score test reviewed in \cite{varin2011overview}.
We focus on Wald-type tests in this paper, but our rationale can be easily extended to CLR and score tests.

\section{Parsimonious composite likelihood testing} \label{sec:method}
\subsection{Optimal Wald composite test under sparse local alternatives} \label{sec:optimal}

Recall that $\delta = \theta_1 -\theta_0$ as defined in Section~\ref{sec:tests} is equivalent to
$\delta(w_\all)=\theta_1(w_\all)-\theta_0(w_\all)$ which is a $p\times\Ncl$ matrix of effective dimension
$q$ giving the group difference. Given a composition rule $w$ which is an $\Ncl$-vector of 1s and 0s,
let $\delta(w)$ be the same as $\delta(w_\all)$ except its $j$th column $\delta_j(w)=0$ whenever $w_j=0$,
$j=1,\dots, \Ncl$. Following the discussion in Section~\ref{sec2.1}, we still use $d_w$ to denote the effective
dimension of $\delta(w)$ knowing that $d_w\leq q$,
and we want to test $H_0: \delta=0$ against $H_1: \delta \neq 0$.
Since some sub-likelihoods for the data of the pooled vector variable $Y=(Y^{(0)T},Y^{(1)T})^T$ may not
contain any significant information about $\delta$, testing these hypotheses using all candidate sub-likelihoods
without selection is unlikely to have a good power.

A plausible approach to overcoming this difficulty is to use more specific alternative hypotheses
by incorporating the composite rule information.
Since models containing redundant sub-likelihoods are unlikely to efficiently capture the group difference
information, we prefer to exclude them from consideration in our test by further adding a
sparsity specification on the composition rule to the alternative hypothesis.
Now we expect a powerful group difference test can be achieved by sequentially testing the
null hypothesis against some alternatives containing a priori composition information and sparsity specification:
\begin{align}
H_0: \delta=  0 \ \ \text{against } \ \ \ H_1: \delta(w) &  \neq 0 \;\; \text{and} \;\;\Vert w \Vert =\Ncl^\ast.
\label{eq3}
\end{align}
Here $w$ is a composition rule given a priori; and $\Vert w \Vert = \Ncl^\ast$, with $\Ncl^\ast\leq \Ncl$ also given a priori,
is regarded as a constraint on the model composition complexity. We will investigate how to choose $w$ and $\Ncl^\ast$
in detail in sections~\ref{sec:alg} and \ref{sec:modelcomplexity}.

Given a composition rule $w$ of size $\Ncl^\ast$, we consider an MCLE of
$\delta$ defined as $\deltahat(w) =\thetahat_1(w) - \thetahat_0(w)$, where $\thetahat_g(w)$, $g=0,1$,
are group-specific profiled MCLEs, and test (\ref{eq3}) by the Wald test statistic
\begin{align}\label{ideal}
T(w)\equiv  n\hat{\delta}(w)^{T} V(w)^{-1} \hat{\delta}(w),
\end{align}
where $V(w)$ is the asymptotic variance matrix of $\sqrt{n}\deltahat(w)$. For given $\Ncl^\ast$,
we assume there is an optimal  composition rule, $w^\ast \in \{0,1\}^{\Ncl}$,
typically with size $\Vert w^\ast \Vert=\Ncl^\ast$ much smaller than $\Ncl$, such that the
corresponding test statistic $T(w^\ast)$ is most powerful among those derived from
all composition rules of size $\Ncl^\ast$. Namely,
\begin{align} \label{Waldo}
w^\ast = \argmax_{w: \ \Vert w \Vert = \Ncl^\ast} \ \ P\left\{ \chi^2(d_w, \lambda(w)) >
Q_\alpha(d_w)  \right\},
\end{align}
where $\chi^2(a,b)$ denotes a random variable following a non-central chi-square distribution with degree of
freedom $a$ and non-centrality parameter $b$, and $Q_\alpha(k)$ is the upper $\alpha$-quantile of
$\chi^{2}(k,0)$ with $\alpha$ being the significance level. The non-centrality quantity in
(\ref{Waldo}) is $\lambda(w) = nE(\hat{\delta}(w))^{T}V(w)^{-1}E(\hat{\delta}(w))$.

The optimal test statistic $T(w^\ast)$ has a straightforward interpretation:
it is determined by the MCLE $\hat{\delta}(w^\ast)$ and according to (\ref{Waldo}), gives the
largest power among all Wald test statistics of form (\ref{ideal}) for testing (\ref{eq3}).
Under $H_0: \delta=0$, $T(w^\ast)$ follows the chi-square distribution with degrees of freedom
 $d_{w^\ast}$ as $n \rightarrow \infty$ when $w^\ast$ is given. This null
distribution is the same as that for the usual Wald test statistic (\ref{Wald}), except that the degree of freedom
$d_{w^\ast}$ may be smaller than $q$ due to the use of the informative composition rule $w^\ast$.

\subsection{Forward-search algorithm} \label{sec:alg}

The ideal test statistic $T(w^\ast)$ outlined in the previous section is appealing from a theoretical viewpoint, but
not very useful in practice, since it is not obvious how to compute the optimal composition rule $w^\ast$ and the
asymptotic covariance $V(w^\ast)$. Such quantities need to be carefully estimated in order to maintain $T(w^\ast)$'s power.
We first proceed to estimate the optimal composition rule $w^\ast$, which is computationally challenging
even when the number of feasible sub-likelihoods $\Ncl$ is moderate since the search space contains
$2^{\Ncl}-1$ possible composition rules. Assuming $\Ncl$ is given and $\widehat{V}(w)$ is available for
estimating $V(w)$, we propose the
following step-up forward search algorithm to efficiently estimate $w^\ast$.

Let $w_{\A} \in \{0,1\}^{\Ncl}$ be a vector with its elements at index $\A \subseteq \{1, \dots, \Ncl\}$ equal to 1,
and zero elsewhere. At each iteration $t=0, 1, 2, \dots$ of the
following algorithm, $\A^{(t)}$ denotes the index set of the active sub-likelihoods used in the Wald test statistic (\ref{ideal}).

\noindent\makebox[\linewidth]{\rule{\textwidth}{0.4pt}}
\label{algorithm1}
\textbf{Main Algorithm: forward step-up composite likelihood (FS-CL) based test}\\
\noindent\makebox[\linewidth]{\rule{\textwidth}{0.4pt}}
\begin{enumerate}
\item[0.]
Initialization. Set $t=0$ (iteration counter) and $\A^{(0)} = \emptyset$
(active set of sub-likelihoods).
\item
Find a new sub-likelihood component with its index
$$
h^{(t+1)} =\underset{i\in \overline{\A}^{(t)}}{\text{argmax}} \  \
n\hat{\delta}(w_i^{(t)})^{T} \{\widehat{V}(w_i^{(t)})\}^{-1}
\hat{\delta}(w_i^{(t)}) \equiv \underset{i\in \overline{\A}^{(t)}}{\text{argmax}}  \ \lambda^{(t)}_i
$$
where $w_i^{(t)}= w_{\A^{(t)} \cup \{i\}}$ augmenting $w_{\A^{(t)}}$,
$\overline{\A}^{(t)}=\{1,\dots, \Ncl\} \setminus \A^{(t)}$ complementing $\A^{(t)}$, $\hat{\delta}(w)$ is the MCLE of
$\delta(w)$ and $\widehat{V}(w)$ is a consistent estimate of $V(w)$.
\item
Update the active set of sub-likelihoods $\A^{(t+1)}=
\A^{(t)}\cup{\{h^{(t+1)}\}}$.
\item
Set $t=t+1$. Repeat 1 and 2 if $t < \Ncl^\ast$. Otherwise, stop the
algorithm and obtain the composition rule $\hat{w}\equiv w_{\A^{(\Ncl^\ast)}}$, regarding it as an optimal estimate of
$w^\ast$.
\end{enumerate}
\noindent\makebox[\linewidth]{\rule{\textwidth}{0.4pt}}

The rationale underlying the above algorithm is similar to well-established step-wise
algorithms used in the context of regression variable selection.
Step 1 finds the most promising sub-likelihood component in terms of its added signal relative to noise
in the current test statistics.
Step 2 simply augments the current active set of sub-likelihoods, $\A^{(t+1)}$, by
including the newly selected sub-likelihood. Step 3 gives a stopping criterion in
terms of the allowed maximum number of sub-likelihood components, $\Ncl^\ast$, which is regarded as a
complexity parameter for the overall composite likelihood function. A separate discussion on the choice of
$\Ncl^\ast$ is given in Section \ref{sec:modelcomplexity}.

The algorithm carries out $\Ncl^\ast(\Ncl-0.5(\Ncl^\ast-1))$ evaluations of the MCLE of $\delta$, which
is much smaller than the exponential rate in exhaustive evaluation.  The final test statistic is
\begin{align}\label{estimated}
T_{\fscl} =  n\hat{\delta}(\hat{w})^{T}
\widehat{V}(\hat{w})^{-1} \hat{\delta}(\hat{w}).
\end{align}
Once the null distribution of (\ref{estimated}) is determined, which will be detailed in Section~\ref{sec:nulldistribution},
$T_{\fscl}$ will be used to test (\ref{eq3}).
Note that if $\Ncl^\ast=\Ncl$, the resulting test is then equivalent to the classic Wald test including all
the sub-likelihood components but may incur much unnecessary computing. However,
the test can be much more powerful if many sub-likelihoods are
redundant and be computationally efficient if $\Ncl^\ast$ is not large.
When $\Ncl^\ast = 1$ and there is only one parameter to estimate in each sub-likelihood component (i.e. $p =1$),
$T_\fscl$ will have the same form as the $T^2_\uminp$ test statistic of \cite{pan2009asymptotic}
given in  Section~\ref{sec:tests}.

It is difficult to estimate the asymptotic covariance matrix $V(w)$ based on the analytical formula provided at
the end of Section~\ref{sec2.1}. Instead, the estimated asymptotic covariance matrix $\widehat{V}(w)$
in Step $1$ of the algorithm can be obtained by using non-parametric bootstrap.
Specifically, we sample with replacement the observations within each group and then compute the bootstrap
replicates of the MCLE for $\delta(w)$, denoted as $\delta^\ast_1(w),\dots,\delta^\ast_B(w)$.
We use the replicates to compute
$\widehat{V}(w)$ as
$$
\widehat{V}(w)=\frac{1}{B-1}\sum^{B}_{b=1}\left(\delta^\ast_b(w)-
\overline{\delta^\ast}(w)\right)\left(\delta^\ast_b(w)-\overline{\delta^\ast}(w)\right)^T,
$$
where $\overline{\delta^\ast}(w)=(1/B)\sum_{b=1}^B \delta^\ast_b(w)$. Our empirical study shows that
the one-dimensional $T_{\fscl}$ statistics is robust under non-parametric bootstrap and it is
sufficient for setting $B=1000$ for most cases in practice. The Jackknife method for computing $\widehat{V}(w)$
may also be used.

\begin{figure}[h]
\centering
\includegraphics[scale=0.7]{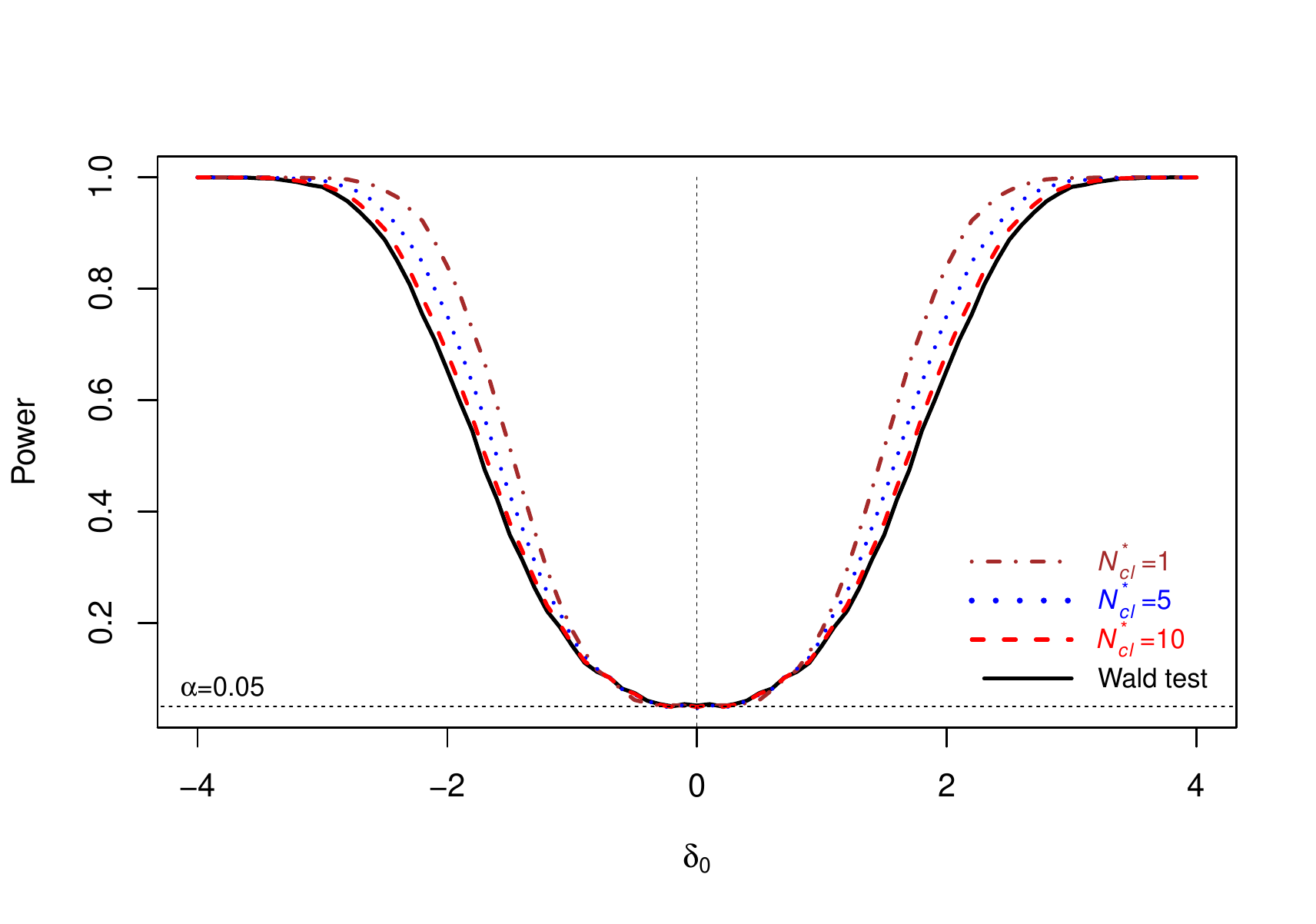}
    \caption{Power of the FS-CL test for hypotheses given in (\ref{eq3}) where the first sample of size 18 come from
$\mathcal{N}_{20}(0,9I)$ and the second sample of size 18 from $\mathcal{N}_{20}(\delta, 9I)$ with
$\delta=(\delta_0, 0,\dots,0)$.
The solid black curve represents the Wald test corresponding to $\Ncl^\ast=20$.
Power curves are estimated using 10,000 Monte Carlo simulations.}
\label{fig:powerplot12}
\end{figure}

Figure~\ref{fig:powerplot12} illustrates the power of the final test statistic $T_{\fscl}$ in a simple simulated example
where two samples of 20-dimensional normal vectors, each of size 18, are generated to test the mean difference.
The normal distribution for the first sample is $\mathcal{N}_{20}(\theta_0\!=\!0,\, 9I)$ and that for the second sample is
$\mathcal{N}_{20}(\theta_1, 9I)$ with $I$ being the $20\times 20$ identity matrix and $\theta_1=(\delta_0, 0,\dots,0)^T$.
Once the data are simulated, we ignore the parameter values underlying the true distributions and proceed to
test $H_0: \delta=\theta_1-\theta_0=0$ at significance level $\alpha=0.05$ using the proposed forward search
algorithm and the FS-CL test statistic $T_{\fscl}$ together with its simulated null distribution.
We set the $\Ncl=20$ marginal sub-likelihoods as all the ones available and consider four specified
values for $\Ncl^\ast$ in $H_1$ in (\ref{eq3}), i.e. $\Ncl^\ast= 1, 5, 10$, or 20 (which gives the classic Wald test).
The power results based on 10,000 simulations of the two-sample data are plotted in Figure~\ref{fig:powerplot12}.
We see that the FS-CL test dominates the Wald
test in terms of power for any $\Ncl^\ast<20$. Clearly the largest power gain is obtained when $\Ncl^\ast=1$,
which should be the case since only the first marginal sub-likelihood contains the information about the
nonzero component in $\delta$ in truth.


\subsection{Null distribution for the FS-CL test}
\label{sec:nulldistribution}

The null distribution of the FS-CL test statistic $T_{\fscl}$ is needed for drawing a conclusion for the test.
Let's first consider a trivial case where the MCLE $\deltahat = (\deltahat_1, \dots,
\deltahat_{\Ncl})$ has its columns independent of each other; and each of its columns has the same
effective dimension $p^\prime$ and the same asymptotic distribution.
Then one can deduce the asymptotic distribution for the FS-CL
test statistic under $H_0: \delta=0$ with given $\Ncl^\ast\leq \Ncl$, which is
\begin{equation}\label{disktppclfs}
T_{\fscl}|\Ncl^\ast \overset{\mathcal{D}}{\rightarrow} \sum_{i=1}^{\Ncl^\ast}
\chi^2_{(i)}(p^\prime)\quad\text{as}\;\;n\rightarrow \infty,
\end{equation}
where ``$\overset{\mathcal{D}}{\rightarrow}$'' stands for convergence in
distribution and $\chi^2_{(1)}(p^\prime)\geq  \cdots \geq \chi^2_{(\Ncl)}(p^\prime)$ are reverse
order statistics from $\Ncl$ independent $\chi^2(p^\prime)$ random variables.
A closed-form expression for the probability density function of
$\sum_{i=1}^{\Ncl^\ast}\chi^2_{(i)}(p^\prime)$ conditional on $\Ncl$ is reported in the Appendix.
From the Bayesian viewpoint, before observing the data there is a
quite large number of equally plausible test statistics  $T_{\fscl}|\Ncl^\ast$, corresponding to a priori
models satisfying the constraint $\Vert w \Vert\leq \Ncl^\ast$. In the Bayesian framework, the complexity
parameter $\Ncl^\ast$ is treated as a random variable with uninformative prior distribution
$\pi(\Ncl^\ast)=(1/\Ncl, \dots, 1/ \Ncl)^T$. Its approximate posterior distribution is discussed in
Section~\ref{sec:modelcomplexity}.

\begin{figure}[ht]
\centering
\includegraphics[scale=0.7]{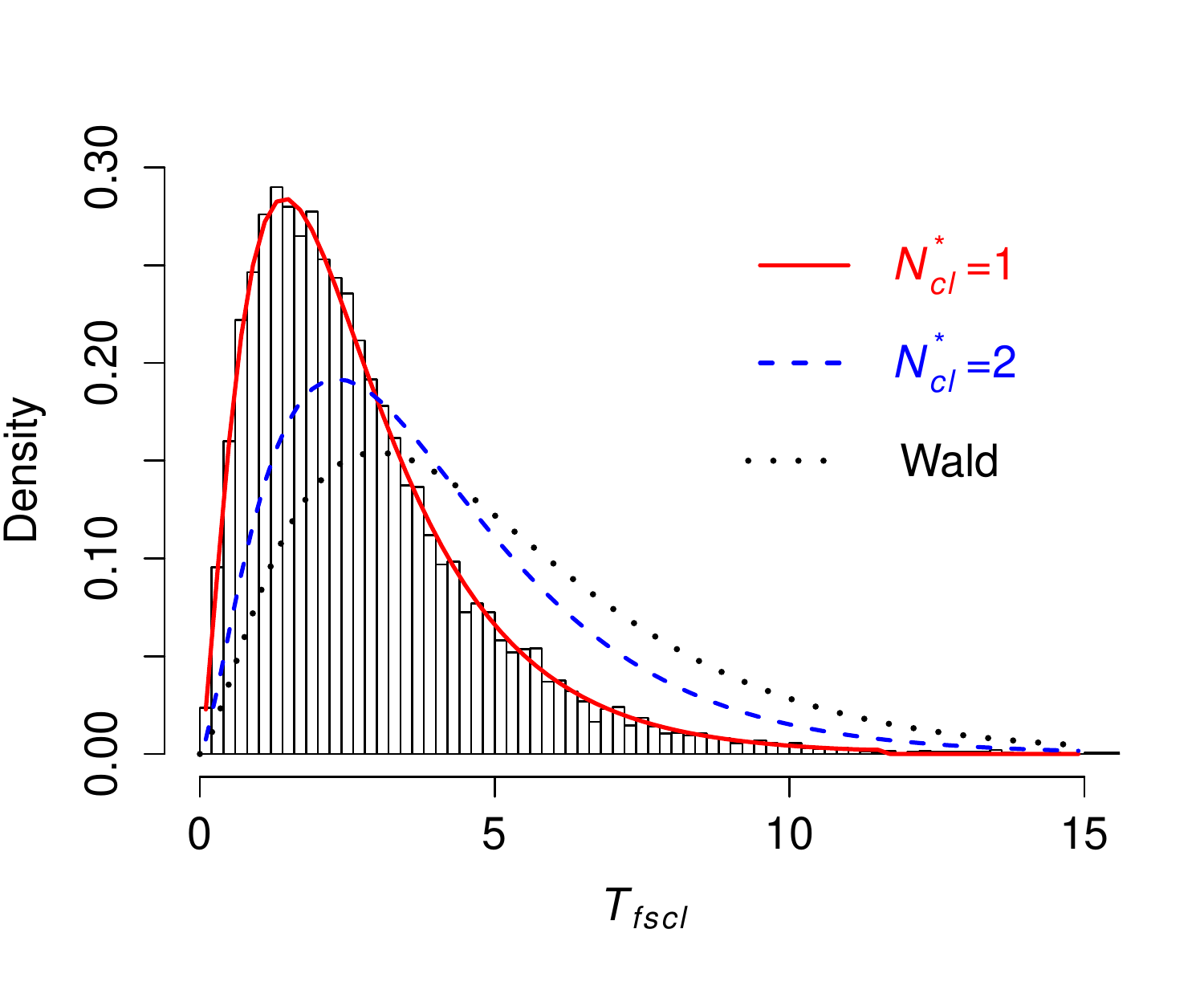}
\caption{Null distribution for the test statistic $T_{\fscl}$ in (\ref{disktppclfs}) when
$p^\prime=5$, $\Ncl^\ast=1,2, 5$, and $\Ncl=5$ (so $\Ncl^\ast=5$ corresponds to the Wald
test). The histogram is generated by Monte Carlo simulation using  $\Ncl^\ast=1$
whereas the smooth pdf curves correspond to the analytical density (\ref{null}) given in the appendix.}
\label{fig:nulldensity}
\end{figure}

Figure~\ref{fig:nulldensity} shows the asymptotic null density $f_T(t|\Ncl^\ast)$ of $T_{\fscl}|\Ncl^\ast$ in
(\ref{disktppclfs}) obtained from formula (\ref{null}) for different values of $\Ncl^\ast$, together with its
histogram estimate at $\Ncl^\ast=1$ obtained from Monte Carlo simulation.
Note that the right tail of the null density becomes lighter when $\Ncl^\ast$ is smaller.
When $\Ncl^\ast$ takes the largest value $\Ncl$ the null density becomes the same as that for the Wald
test statistic (\ref{Wald}).  This property of the null distribution makes it more likely for the FS-CL test than
the classic Wald test to reject the null hypothesis $H_0:\delta=0$ when the alternative hypothesis is true.

In general the columns of the MCLE $\deltahat$ are correlated with each other, thus the null distribution
of $T_{\fscl}$ is difficult to obtain. We propose to use a random permutation method to acquire the null
distribution of the FS-CL test statistic. The main idea is to permute the data many times and use each
permutation to compute a replicate of the test statistic. The empirical distribution of the permutation
replicates is used as an estimated null distribution of the test statistic.

Specifically, we draw all the observations together and randomly distribute them into different groups
with the sample size in each group unchanged. By doing so, each permutation can
be treated as generating a new data sample under the null hypothesis that there are no characteristic differences
between groups. Using each newly generated data sample, we compute the MCLE of the difference parameter $\delta$
and the corresponding FS-CL test statistic as a permutation replicate.
Repeating this procedure for $B$ times, we will then acquire $B$ permutation replicates for the
test statistic, denoted $(T^\ast_{(1)}, \dots, T^\ast_{(B)})$. We use the empirical distribution of
$(T^\ast_{(1)}, \dots, T^\ast_{(B)})$ as an estimate of the null distribution, and use the upper
$\alpha$-quantile as the rejection threshold of the FS-CL test.

\subsection{Choice of $\Ncl^\ast$ and the maximum posterior test statistic}
\label{sec:modelcomplexity}

Note that choosing different $\Ncl^\ast$, the maximum number of
allowed sub-likelihoods in the FS-CL algorithm in Section~\ref{sec:alg}, leads to
different test statistics $T_{\fscl}$ defined in (\ref{estimated}). It is thus
important to discuss how to choose an appropriate value of $\Ncl^\ast$. Since $\Ncl^\ast$ can be regarded as
a model complexity parameter, it seems natural to use a well-established
model-selection criterion for choosing $\Ncl^\ast$.  We propose to use the composite likelihood
Bayesian information criterion  (CL-BIC) studied by \cite{gao2010composite}.
The CL-BIC is a robust generalization of the classic Bayesian information criterion (BIC) not requiring that the
estimating equation used corresponds to the true model. For a
single group of data of sample size $n$, the classic CL-BIC for a composition model including all
sub-likelihoods available is defined by
\begin{align}\label{CL-BIC}
\text{CL-BIC}^\ast = -2\ell_{cl}(\thetahat_{all};w_{all})+ \log(n)\hat{p}^\ast,
\end{align}
where $\ell_{cl}(\theta;w)$ is the weighted composite likelihood function
defined in (\ref{comp_lik}) with $w=(1,\dots,1)^T$, and $\thetahat_{all}$ is the corresponding
MCLE. The term
$\hat{p}^\ast=\text{Tr}(\widehat{H}^{-1}\widehat{J})$ represents the
estimated effective degrees freedom in the parameter with $\text{Tr}(\cdot)$  denoting the trace function, and
$\widehat{H}$ and $\widehat{J}$ are estimates of the Hessian and
score variance obtained as
\begin{align} \label{matrices}
   \widehat{H}\equiv\widehat{H}(w_{\all})= -  \nabla^2
\ell_{cl}(\thetahat_{\all}; w_{\all}) ,  \ \
   \widehat{J} \equiv \widehat{J}(w_{\all})= \widehat{\text{Var}}\left\{ \nabla
\ell_{cl}(\thetahat_{\all}; w_{\all})\right\}.
\end{align}
CL-BIC can naturally be extended for any composition model specified by the composition rule $w$,
where we just need to replace $\hat{\theta}_{\all}$ with $\hat{\theta}_w$, $w_{\all}$ with $w$, and
$\hat{p}^\ast$ with $\hat{p}^\ast_w=\text{Tr}(\widehat{H}^{-1}(w)\widehat{J}(w))$.
In the special case of $H(\theta, w)$ and $J(\theta, w)$ defined in Section~\ref{sec2.1} being equal to
each other (cf. \cite{Lindsay2011issues}), $\hat{p}^\ast_w$ should equal $d_w$ approximately.

In our two-group sequential test,
the group-specific composite likelihoods are
\begin{align} \label{combined}
\ell^{(j)}_{cl}(\theta_j; w) =   \sumkK \dfrac{w_k
\ell^{(j)}_k(\theta_j)}{\sumkK w_k}, \  \ j =0,1,
\end{align}
where $\ell^{(j)}_k$ denotes the $k$th sub-likelihood in group $j$. Thus, we
propose to construct the combined two-sample CL-BIC for a composition rule $w$ as
\begin{align}
\label{TIC}
\text{CL-BIC}(w)=- 2\left\{ \ell^{(0)}_{cl}(\hat{\theta}_0; w) +
\ell^{(1)}_{cl}(\hat{\theta}_1; w) \right\}+ \log(n) \left\{ \hat{p}^{\ast (0)}_w +
\hat{p}^{\ast (1)}_w \right\}
\end{align}
where $\thetahat_0$ and $\thetahat_1$ are group-specific MCLEs, and
$\hat{p}^{\ast (j)}_w=\text{Tr}(\{{\widehat{H}^{(j)}}(w)\}^{-1}\widehat{J}^{(j)} (w))$ with
$\widehat{H}^{(j)}(w)$, $\widehat{J}^{(j)}(w)$ computed similarly as in (\ref{matrices}).
In the special case of $H^{(j)}(\theta_j, w)$ and $J^{(j)}(\theta_j, w)$ being equal to
each other, $j=0,1$, we should have  $\hat{p}^{\ast(0)}_w + \hat{p}^{\ast(1)}_w= 2d_w$ approximately.

With the proposed $\text{CL-BIC}(w)$ we choose the best value of $\Ncl^\ast$ as
$$\Ncl^\ast = \text{argmin}_{1\leq t \leq \Ncl} \text{CL-BIC}(\hat{w}^{(t)}),$$
where $\hat{w}^{(t)}$ is the
final composition rule obtained from the FS-CL Algorithm in Section~\ref{algorithm1} after $t$ steps.

In our setting, the CL-BIC model-selection framework offers a natural interpretation as being induced from
a posterior distribution for the composition complexity parameter $\Ncl^\ast$.
Specifically, under the discrete uniform prior for $\Ncl^\ast$,
$\pi_{\Ncl^\ast}(k) = 1/\Ncl$, $k=1,\cdots, \Ncl$, the posterior of $\Ncl^\ast$ is
$$
\pi_{\Ncl^\ast}(k|Y_1, \dots, Y_n) = \dfrac{\exp \left\{ -\text{CL-BIC}(\hat{w}^{(k)})\right\} }{\sum_{s=1}^{\Ncl}
\exp\left\{-\text{CL-BIC}(\hat{w}^{(s)})\right\}}.
$$
Then, the best $\Ncl^\ast$ value corresponds to the maximum a posteriori (MAP) estimate of $\Ncl^\ast$
and the resulting test statistic $T_{\fscl}|\Ncl^\ast$ is referred to
as maximum a posteriori test statistic.

\section{Numerical examples and simulation study} \label{MonteCarlo}
\label{sec:numerical}

\textbf{Example 1: Normally distributed MCLEs.} Consider a simulated example of MCLE-based testing for
the difference of group means from two samples of 40-dimensional normal data with known covariance matrix.
This is equivalent to testing $H_{0}: \sqrt{n}\deltahat \sim \mathcal{N}_{40}(0 ,V)$ against
$H_{1}: \sqrt{n}\deltahat \sim\mathcal{N}_{40}(\delta,V)$,
where $\deltahat$ is the MCLE of $\delta$ having $40$ elements.
We set the considered composite likelihood function to comprise up to $\Ncl=20$ independent pairwise sub-likelihoods,
where each candidate sub-likelihood is for a 2-dimensional subset of the data and contains just two elements of $\delta$.
In generating the data satisfying $H_1$, we consider four models ($m=1,\dots,4$) and set the elements of $\delta$ as
$\delta_{j}=(-1)^j\cdot0.5(m+1)\cdot I(j\leq 6-m)$, $j=1,\dots,40$, where $I(\cdot)$ is the indicator function.
 The covariance matrix of $\sqrt{n}\deltahat$ is set as
\begin{align*}
V= I_{\Ncl} \otimes
   \begin{pmatrix}
   1.5  &  0.2 \\   0.2 & 1
   \end{pmatrix},
\end{align*}
where $I_{\Ncl}$ is an $\Ncl$-dimensional identity matrix and ``$\otimes$" denotes the Kronecker product.
With this setting the number of the pairwise sub-likelihoods containing nonzero elements of $\delta$ under $H_1$
decreases from 3 to 1 as $m$ increases from 1 to 4, while the magnitude of each nonzero $\delta_j$ increases.
Using the above setting we generate 10,000 replicates of $\sqrt{n}\deltahat$ under $H_0$ and $H_1$,
respectively. After that we set the significance level $\alpha=0.05$, and compute the Monte Carlo estimates
of the Type I error probability and the power for the FS-CL test, the Wald test and two of its variants LSSB and LSSBw.
The results are summarized in Table~\ref{Table1}.

\begin{table}[htp]
\centering
\begin{tabular}{ cccccccccc }
\hline
Model ($m$) & \multicolumn{4}{c}{Type I error}&& \multicolumn{4}{c}{Power} \\
             & FS-CL & Wald & LSSB & LSSBw& & FS-CL & Wald & LSSB & LSSBw   \\ \hline
1           & 0.0461 & 0.0447 & 0.0462 & 0.0467 && 0.2736& 0.2785 & 0.2040 & 0.2159   \\
2           & 0.0481 & 0.0466 & 0.0467 & 0.0458 && 0.5793 & 0.5436 & 0.4073 & 0.4426  \\
3           & 0.0491 & 0.0499 & 0.0501 & 0.0495 && 0.7951 & 0.7153 & 0.5615 & 0.6015  \\
4           & 0.0509 & 0.0506 & 0.0511 & 0.0527 && 0.8568 & 0.7423 & 0.5977 & 0.6355  \\ \hline
\end{tabular}
\caption{Estimated Type I error and power for the FS-CL test, the Wald test, and the LSSB and LSSBw tests
described in Section \ref{sec:tests}, for testing $H_0$ vs. $H_1$. Results are based on $10,000$ samples
generated according to the setting described in Example 1.}
\label{Table1}
\end{table}

From Table \ref{Table1}, we see all the tests have similar Type I error probabilities around $\alpha$.
However, although the LSSB and LSSBw
tests are computationally easier than our FS-CL test, they are inferior in terms of power. This is expected since the LSSB
test sets $V=I$ and the LSSBw test uses only the diagonal elements of $V$, leading to loss of information on the
correlation between components of $\deltahat$ belonging to the same sub-likelihood.
In summary, the FS-CL test performs uniformly better than all the other tests in terms of power,
in all situations where the number of informative sub-likelihoods involved is sparse.

Finally, we assessed the quality of the model-selection procedure annexed to our FS-CL algorithm by
computing the Monte Carlo estimates of the Hamming distance between the optimally estimated composition rule
$\hat{w}$ given by the FS-CL test and the true composition rule $w^\ast$ informed by each of the prescribed
models. In all our simulations we have found that the Hamming distance is essentially equal to 0 up to a negligible
simulation error, which confirms the FS-CL test systematically selects
only informative sub-likelihoods to construct the test statistic $T_\fscl$.

\paragraph{Example 2: Latent multivariate Gaussian model.} Now we investigate the performance of our new test
for categorical data analysis based on a latent multivariate Gaussian model. This model has been previously
studied for analyzing single nucleotide polymorphisms (SNPs) data (cf. \cite{pan2009asymptotic, han2012composite}).
The advantage of using the latent multivariate Gaussian model is its ability to model correlated categorical
variables through the latent quantiles and covariance matrix.

Consider independent $d$-vector observations $Y_i =(Y_{i1}, \dots, Y_{id})^T$, $i=1,\dots,n$, with each
element $Y_{ij}$ being categorical taking one of $C$ labels $1, \dots, C$.
For the $i$th $d$-vector observation, we assume there is a latent vector variable
$Z_{i}=(Z_{i1}, \dots,Z_{id})^T$ that follows a multivariate Gaussian distribution,
$Z_{i}\sim \mathcal{N}_d(0,\Sigma)$,  where $\Sigma$ is a $d\times d$ correlation matrix.
We also assume the existence of $C-1$ quantile constants $\gamma_{j1}, \gamma_{j2}, \dots, \gamma_{j(C-1)}$
for each $j=1,\dots, d$, such that for all $i=1,\dots, n$, $Y_{ij}$ takes label $k$ if $Z_{ij}\in \Gamma_{jk}\equiv
(\gamma_{j(k-1)}, \, \gamma_{jk}]$, where $k=1,\dots, C$ with $\gamma_{j0}=-\infty$ and $\gamma_{jC}=\infty$.
It is easy to see that the marginal and joint probability distributions of $(Y_{i1}, \dots, Y_{id})$ are
determined by the quantile parameters $\gamma=\{\gamma_{j1}, \cdots, \gamma_{j(C-1)};\, j=1,\cdots,d\}$
and correlation matrix parameter $\Sigma$. For example, $\Pr(Y_{ij}=2) =\Pr(\gamma_{j1}<Z_{ij}\leq \gamma_{j2})$
and $\Pr(Y_{ij}=2, Y_{ij^\prime}=3) =\Pr(\gamma_{j1}<Z_{ij}\leq \gamma_{j2}, \,
\gamma_{j^\prime 2}<Z_{ij^\prime}\leq \gamma_{j^\prime 3})$, etc.
We use $\theta$ to denote the vector collecting the elements in $\gamma$ and $\Sigma$.

Let  $f(z_1, \dots, z_d;\Sigma)$ denote the $d$-variate normal density $\mathcal{N}_d(0,\Sigma)$.
Now estimating the quantile parameters $\gamma$ from the data $(Y_1,\cdots, Y_n)$ can be done by
maximizing the weighted one-wise marginal composite likelihood function
\begin{align} \label{CL1}
&CL_{1}(\gamma)= \prod_{i=1}^{n}\prod_{j=1}^{d}\left(\prod_{k=1}^C\left[\int_{\Gamma_{jk}}
f(z_{j};1)d z_{j}\right]^{I(Y_{ij}=k)}\right)^{w_j},
\end{align}
where $I(Y_{ij}=k)$ is an indicator function and $(w_1,\cdots, w_d)$ is the binary weight vector (hence
$\Ncl=d$ here). One can extend (\ref{CL1}) by including pairwise likelihood components as
described in \cite{han2012composite} so that $\gamma$
and $\Sigma$ can be estimated simultaneously.

We consider having two groups (case and control) of $d=6$ dimensional observations each taking
one of $C=3$ labels (categories). The correlation structure of the latent vector variable is set to be
$\Sigma=I$ where $I$ is a $d$-dimensional  identity matrix. The group-specific latent parameters
are then $\theta_0=\gamma_0$ and
$\theta_1=\gamma_1$.  For the control group we set
$\gamma_{0}=(-0.3,0.3;-0.3,0.3;-0.3,0.3;-0.3,0.3;-0.3,0.3;-0.3,0.3)$ as the true values, while for the
case group we set $\gamma_1 = \gamma_0 + \delta$ where
$\delta=(-\epsilon,\epsilon;0,0;0,0;0,0;0,0;0,0)$ and $\epsilon  = 0.3, 0.4, 0.5$. We
generate 1000 Monte Carlo samples of size $n= 200$  ($100$ controls and $100$ cases) using the above
setting, and denote the group-specific MCLEs by
$\hat{\gamma}_{0}$ and $\hat{\gamma}_{1}$, respectively, where $\hat{\gamma}_{0}$ and $\hat{\gamma}_{1}$
are $2\times6$ dimensional vectors. For each sample, we compute the MCLE difference
$\hat{\delta}  = \hat{\gamma}_{1}-\hat{\gamma}_{0}$ and estimate the asymptotic covariance matrix
$V$ of $\sqrt{n}\hat{\delta}$ by nonparametric bootstrap as described in Section \ref{sec:alg}.
We then perform the FS-CL test and the Wald, LSSB and LSSBw tests discussed in the paper for testing $H_0: \delta=0$
against a sparse local alternative. We use the permutation method to simulate the null distributions and the
associated 0.05 level critical values in these tests, by which we compute the Monte Carlo estimates
of the Type I error and the power of these tests based on the 1000 generated samples.

\begin{table}[h]
\centering
\scalebox{1}{
\begin{tabular}{ cccccccccc } \hline
            &  \multicolumn{5}{c}{FS-CL}             & Wald & LSSB & LSSBw  \\
$\epsilon$ $\setminus$ $\Ncl^\ast$  & 1& 2 & 3 & 4&  5 &6 &  &  \\
\hline
$0$         &  0.045 & 0.052 & 0.048 & 0.047 & 0.046 & 0.045&  0.056 & 0.053         \\
$0.3$       &  0.614 & 0.585 & 0.543 & 0.518 & 0.501 & 0.496&  0.203 & 0.181        \\
$0.4$       &  0.873 & 0.840 & 0.812 & 0.776 & 0.755 & 0.752& 0.380 & 0.328        \\
$0.5$       &  0.979 & 0.972 & 0.950 & 0.938  & 0.927 & 0.924& 0.653 & 0.584 \\ \hline
\end{tabular}}
\caption{Monte Carlo estimates of Type I error probability ($\epsilon=0$) and power ($\epsilon>0$) of the
various tests when the data have the latent multivariate Gaussian model described in
Example 2. The tests considered are FS-CL with $\Ncl^\ast$ ranging from 1 to 6 and the Wald test, LSSB and LSSBw
tests described in Section \ref{sec:framework}. Note that the Wald test corresponds to $\Ncl^\ast=6$ (no selection).
Results are based on $1000$ Monte Carlo samples.}
\label{Table2}
\end{table}

Table~\ref{Table2} gives the Monte Carlo estimates of the Type I error probability (in row $\epsilon = 0$) and the
power (in rows $\epsilon > 0$). The table reveals the power for the FS-CL test is considerably larger than that
for all the other tests in all simulated situations of group difference of size $\epsilon$.
Specifically, the power improvement is dramatic when comparing the FS-CL test with the LSSB and LSSBw tests
for values of $\epsilon$ closer to zero.

Also Table~\ref{Table3} shows the values of CL-BIC described in Section~\ref{sec:modelcomplexity}
for each $\Ncl^\ast$. We see CL-BIC is minimized at $\Ncl^\ast=1$ in all the scenarios of $\epsilon$.
This result conforms to the true setting used in generating the Monte Carlo samples. Namely, only the first
one-wise marginal sub-likelihood contains the information about $\epsilon$.
Thus the power of the FS-CL test should be the largest at selecting $\Ncl^\ast=1$, which is clearly confirmed by
the results in Table~\ref{Table2}.

\begin{table}[h]
\centering
\scalebox{1}{
\begin{tabular}{ cccccccc }
\hline
   $\epsilon$  $\setminus$ $\Ncl^\ast$        & 1   & 2  & 3 & 4 &  5     &        6            \\
\hline
$0.3$     &  431.76 & 440.49 & 449.11 & 457.46& 465.90& 474.23       \\
$0.4$     & 424.61 & 436.60 & 446.48& 455.46& 464.29& 472.86  \\
$0.5$     &  414.31 & 431.38 & 442.96& 452.83& 462.19& 471.13 \\ \hline
\end{tabular}}
\caption{Composite likelihood Bayesian information criterion (CL-BIC) values at different maximum numbers of steps
($\Ncl^\ast$) in the forward step-up algorithm described in Section~\ref{sec:alg}.
The values are the averages computed based on 1000 Monte Carlo samples generated from the multivariate latent
Gaussian model in Example~2 with various magnitudes of the parameter difference under the alternative
hypothesis (i.e. $\epsilon=0.3,0.4,0.5$). }
\label{Table3}
\end{table}

\noindent\textbf{Example 3: Effects of increasing the number of candidate sub-likelihoods.}
We continue by considering the latent Gaussian model underlying the case-control data  described
in Example 2  to see how the FS-CL test procedure and the other discussed tests perform as  the number
of candidate one-wise sub-likelihoods, $\Ncl$, grows. In the set-up we let $\Ncl=d$ change from 6 to 20 but
let only the first one-wise sub-likelihood contain the information of case-control difference.
We continue to assume $C=3$ categories for each variable in the data. But differently from Example 2, we consider
parameter vectors $\gamma_0^{(m)}$ and $\gamma_1^{(m)}$  of
length $2(m+5)$, for $m=1, \dots 15$ (note $\Ncl=m+1$). Specifically, we set
$\gamma^{(m)}_{0,i}=0.3\times(-1)^i$, $i=1,\cdots, 2(m+5)$,
and $\gamma^{(m)}_{1} =\gamma^{(m)}_{0}+\delta^{(m)}$
where $\delta^{(m)}$ is a vector of length $2(m+5)$ with $i$th element $\delta_i=0.8\times(-1)^i$,
if $i\leq2$, and $\delta_i=0$ otherwise.
The covariance matrix $\Sigma$ of the latent vector variable is set as the identity matrix.
For each model $m=1,\dots, 15$, we generate 200 Monte Carlo samples of size $n=120$ (60 cases and 60 controls)
and estimate the test power where the significance level is set as $\alpha=0.05$.

\begin{figure}[t]
     \centering
     \includegraphics[scale=0.7]{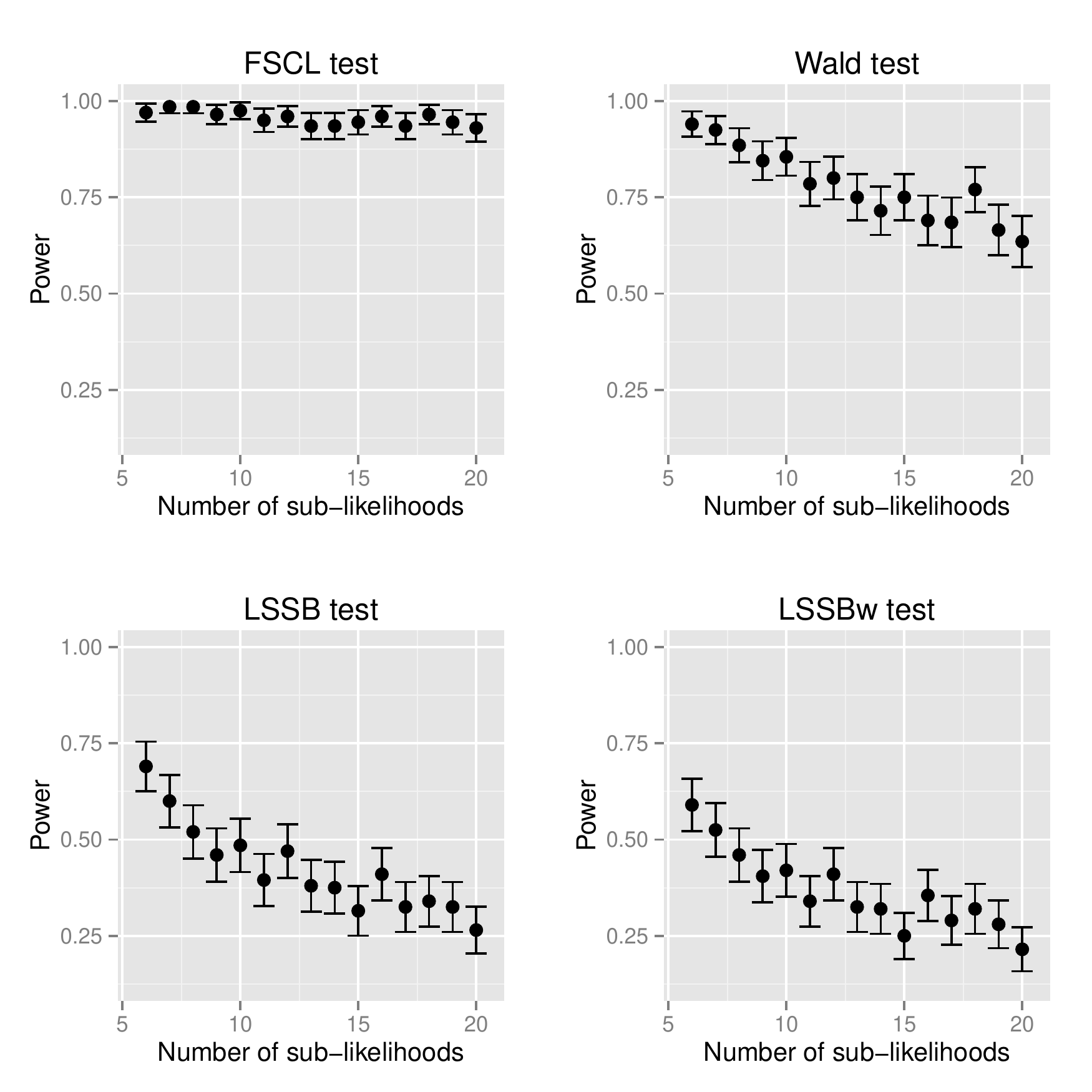}
          \caption{Monte Carlo estimates of the power for the FS-CL test, and the Wald, LSSB and LSSBw tests
described in Section \ref{sec:tests} for increasing numbers of sub-likelihoods ranging from 6 to 20.
Vertical bars denote the simulated 95\% probability intervals of the power. The results are
obtained based on 200 Monte Carlo samples from the multivariate latent Gaussian model as described in Example 3.}
     \label{fig:snpchangepic}
\end{figure}

Figure~\ref{fig:snpchangepic} displays the Monte Carlo estimates of power (represented by the dots),
together with their 95\% probability intervals, for the FS-CL, Wald, LSSB and LSSBw tests as the number of
candidate sub-likelihoods grows. As the number of uninformative sub-likelihoods increases,
it shows the LSSB  and LSSBw tests are increasingly weak compared to the FS-CL and
Wald tests. The FS-CL and Wald tests have similar power when the number of candidate sub-likelihoods is small.
Remarkably, the Wald test's performance decreases dramatically when $\Ncl$ increases, while the power of the
FS-CL test remains stable regardless of the number of irrelevant sub-likelihoods considered.
This behavior can be explained by noting that, as the data dimension $d$ (consequently the number of candidate
sub-likelihoods $\Ncl$) increases, more noise is added to the unweighted composite likelihood.
Therefore those sub-likelihoods informative for distinguishing the alternative hypothesis from the null
will become less significant in the Wald, LSSB  and LSSBw tests.
In contrast, the FS-CL test tends to keep such informative sub-likelihoods and to remove the noisy ones,
therefore having achieved a stable high power (always near $0.9$ in Figure~\ref{fig:snpchangepic}).

\subsection{Analysis of the Australian Breast Cancer Family genomic data}
\label{sec:real}
In this section, we apply the FS-CL procedure and the Wald, LSSB and LSSBw tests to data from
a case-control study on breast cancer. Cases are obtained from the Australian Breast Cancer Family
(ABCF) study \citep{mccredie2003risk} while controls are from the Australian Mammographic Density Twins and
Sisters Study \citep{odefrey2010common}. The data set consists of 356 observations (284 controls and
72 cases) on 100 SNPs. SNPs are the mutated pairs of single nucleotide
(A,T,C,G) in a DNA sequence. These mutated pairs can be categorized into three groups denoted as 0, 1 and 2
(0 and 2 are homozygous and 1 denotes heterozygous).
After recommended data cleaning and quality control, the final dataset comprises
356 vector observations on 61 SNP variables and contains no missing data.
Our objective is to test the significance of association between the SNPs and breast cancer.

\begin{figure}[ht]
    \centering
    \includegraphics[scale=0.111]{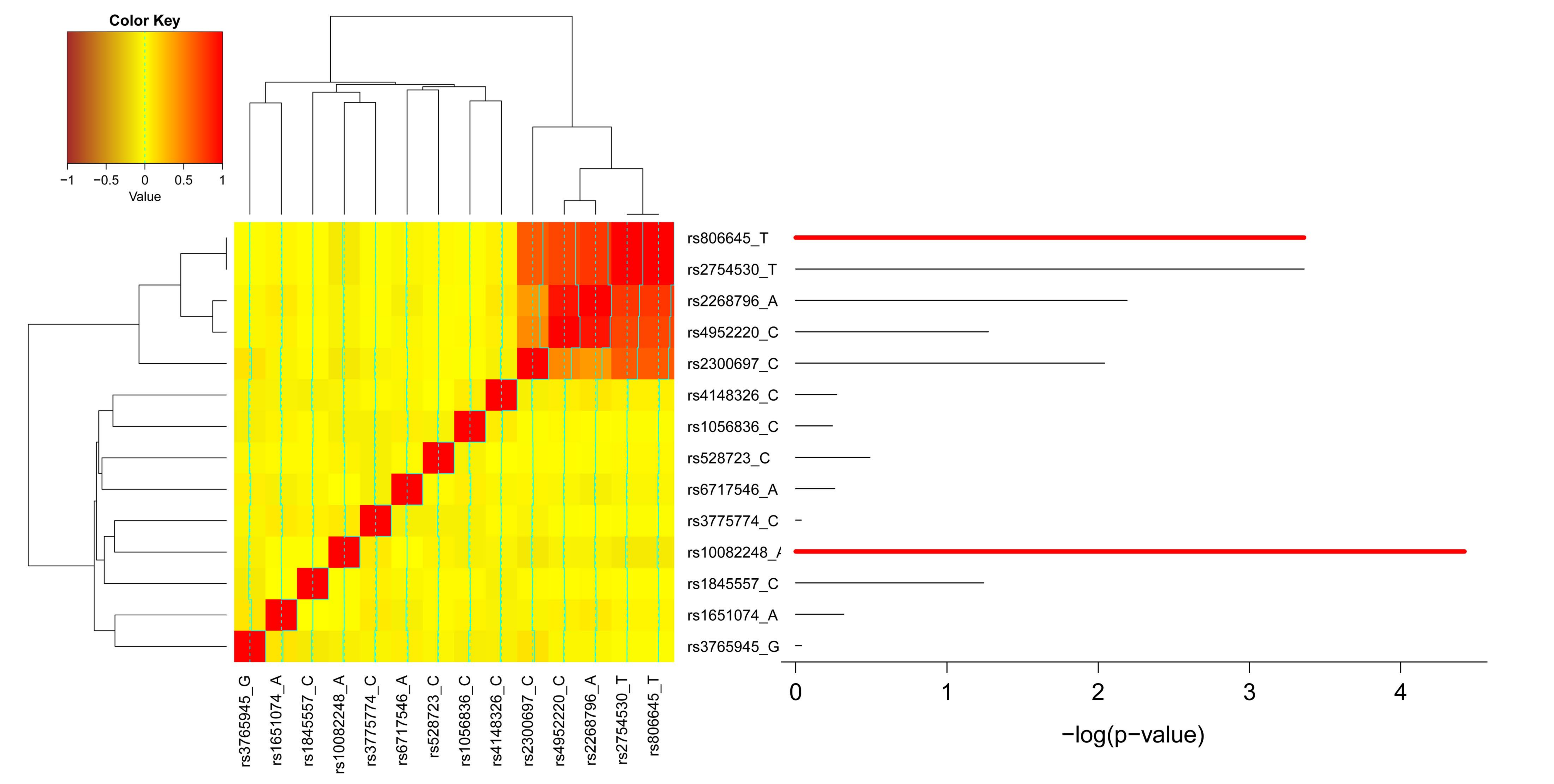}
    \caption{SNP plot of the 14 SNPs used in assessing the group difference in regard to their associations with
breast cancer. The first SNP and the last 9 SNPs are used in Case 1 in Section~\ref{sec:real}; and the first 5 SNPs
are used in Case 2 in Section~\ref{sec:real}. The right hand side of the figure shows the $p$-values (under a negative log scale)
of the association tests between individual SNPs and breast cancer. The thick lines refer to the two selected SNPs by
the FS-CL procedure. LHS of the figure shows the correlation heat map among SNPs (light color for small correlation). }
    \label{fig:snpplot1}
\end{figure}

\textit{Case 1: Weakly dependent SNPs.}  In order to illustrate our testing procedure in the context of weakly
dependent SNPs, we select 10 SNPs (\texttt{rs10082248\_A}, \texttt{rs806645\_T}, \texttt{rs3765945\_G},
\texttt{rs1056836\_C}, \texttt{rs4148326\_C}, \texttt{rs6717546\_A}, \texttt{rs1845557\_C},
\texttt{rs3775774\_C}, \texttt{rs1651074\_A}, and \texttt{rs528723\_C}) as reported in Figure~\ref{fig:snpplot1}.
These SNPs are selected by investigating sample correlations of all the 61 SNPs and picking those SNPs
with the pairwise sample correlations, among the selected, being smaller than 0.1. To test the significance
of association between the selected SNPs and breast cancer, we fit a latent Gaussian model described in
Example~2 for these SNPs using the maximum composite likelihood method, and then implement the FS-CL
procedure, with various choices of $\Ncl^\ast$, for testing the case-control difference between the
quantile parameters involved in the latent model.

\begin{table}[h]
\centering
\scalebox{0.97}{
\begin{tabular}{ cccccccccccccc}\hline
& \multicolumn{9}{c}{FS-CL}  & Wald  & LSSB & LSSBw \\
$\Ncl^\ast$    & 1    & 2  & 3 & 4 &  5     &        6 & 7& 8 &   9  &  10   &     &        \\
\hline
CL-BIC         &  676 & 465 &  573 &  651  & 653  & 703 & 750 &782 & 821 &   \\
$p$-value    &0.12  &0.04&  0.08&  0.08 & 0.09 & 0.10 & 0.10  & 0.10 & 0.09  & 0.09    &  0.50  & 0.27 \\ \hline
\end{tabular}}
\caption{Composite likelihood Bayesian information criterion (CL-BIC) with respect to $\Ncl^\ast$ ranking
from 1 to 9, as well as the $p$-values of the FS-CL test and the Wald type tests described in Section~\ref{sec:tests},
for the 10 weakly dependent SNPs. The CL-BIC values are computed using (\ref{TIC}), and the $p$-values of
the FS-CL test are acquired from the permutation null distribution of the test statistics as described in
Section~\ref{sec:nulldistribution}. }
\label{Table4}
\end{table}

Table~\ref{Table4} shows the $p$-values of the FS-CL test, as well as the CL-BIC values for $\Ncl^\ast$
ranking from 1 to 9. The CL-BIC values suggest that $\Ncl^\ast=2$ yields the best fitted model.
When $\Ncl^\ast=2$, the $p$-value of the FS-CL test is 0.04, while the $p$-value for the Wald, LSSB
and LSSBw tests are 0.09, 0.50, and 0.27 respectively. At the 0.05 significance level the FS-CL test rejects the
null hypothesis of no SNPs association with the disease, while the other tests cannot reach the same conclusion.

When $\Ncl^\ast=2$, the FS-CL procedure selects SNPs \texttt{rs806645\_T} and \texttt{rs10082248\_A} as having
significant association with the disease. To investigate the validity of this selection, we conduct marginal
chi-square association tests for between individual SNPs and the disease. Figure~\ref{fig:snpplot1} shows the
correlations between the SNPs under consideration and the $p$-values from the marginal association tests.
From Figure~\ref{fig:snpplot1} we see the 10 weakly dependent SNPs considered in this case include the first
SNP from the first 5 correlated ones and the last 9 ones. The two SNPs selected by the FS-CL
have very small $p$-values (marked with thick lines), compared to the other SNPs.

\begin{figure}[ht]
    \centering
    \includegraphics[scale=0.495]{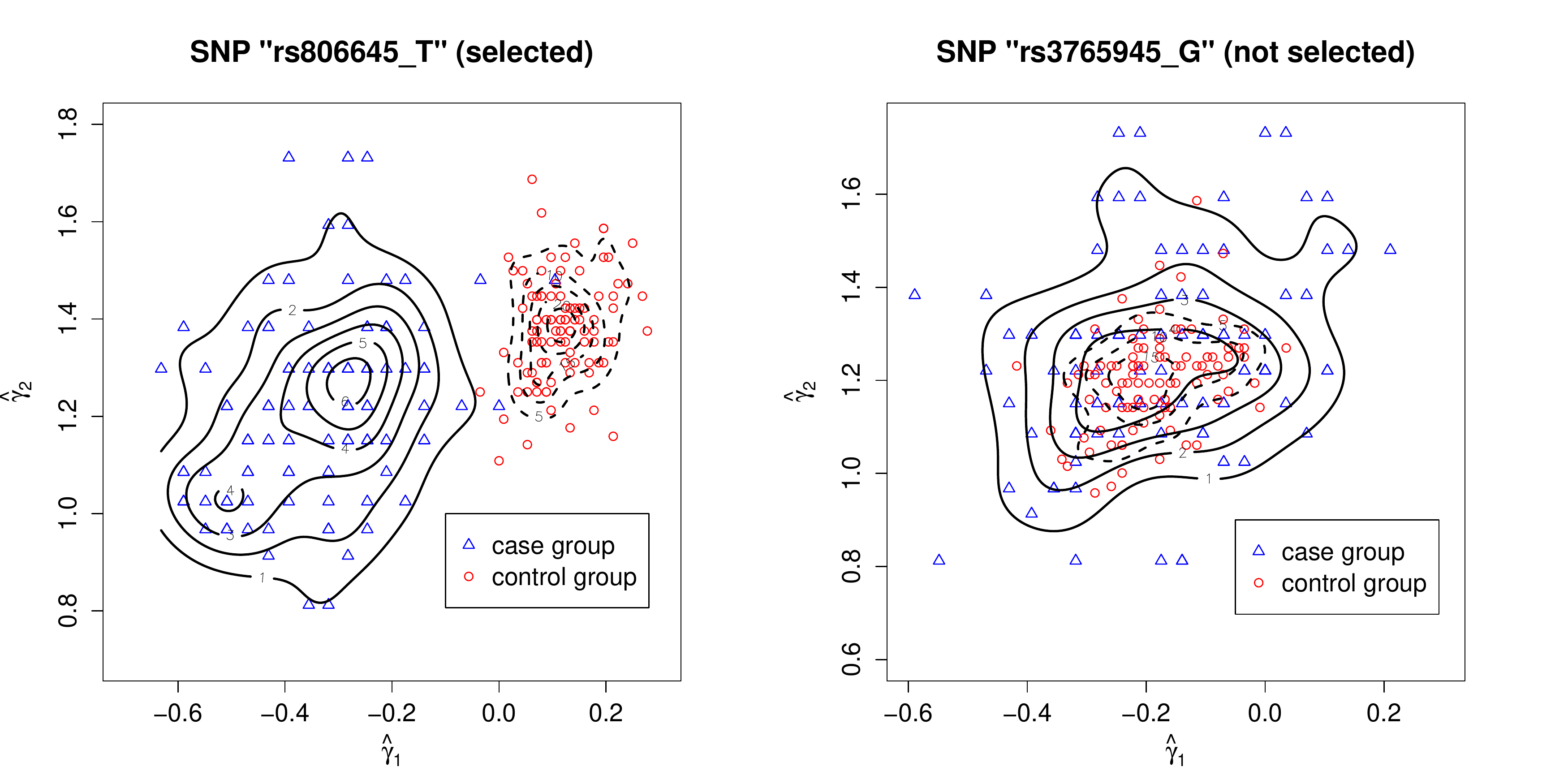}
    \caption{Bootstrap distributions for the quantile parameters estimates $\hat{\gamma}_1$ and $\hat{\gamma}_2$
for the selected SNP \texttt{rs806645\_T} (left) and the unselected SNPs \texttt{rs3765945\_G} (right),
obtained from 100 bootstrap replicates. Triangles represent estimates from the case group and the solid contour lines
specify the estimated confidence regions in the case group. Small circles represent estimates from the control group and the
dashed contour lines specify the estimated confidence regions in the control group.}
    \label{fig:tourplot}
\end{figure}

For illustration, Figure~\ref{fig:tourplot} (left) displays bootstrap distributions of the MCLEs of the latent quantile
parameters for a selected SNP (\texttt{rs806645\_T}) in the respective case and control groups, while
Figure~\ref{fig:tourplot} (right) displays the counterparts for an unselected SNP (\texttt{rs3765945\_G}). The
triangles and circles represent the bootstrap replicates from the case and control groups respectively.
When comparing case and control groups for a selected SNP, the figure implies the quantile parameters values
for the two groups are well separated, concentrating into different clusters.
On the other hand, the bootstrap distributions for an unselected SNP are overlapping and not clearly
distinguishable. Both Figures~\ref{fig:snpplot1} and \ref{fig:tourplot} suggest that the SNPs selected by the FS-CL
procedure are more likely to change their values from control to case,
and they appear to have significant effects on breast cancer.

\textit{Case 2: Dependent SNPs.} Next, we focus on clusters of dependent SNPs having high correlations.
For illustration purpose, we choose the cluster of SNPs \texttt{rs806645\_T}, \texttt{rs2754530\_T}, \texttt{rs2268796\_A},
\texttt{rs4952220\_C}, and \texttt{rs2300697\_C}. They are the first five SNPs in Figure~\ref{fig:snpplot1}
which are highly correlated. Other clusters can also be analyzed which will not be detailed here.

Table~\ref{Table5} shows the $p$-values of the FS-CL test at specified $\Ncl^\ast=1,\dots,4$. It also gives the
corresponding CL-BIC values, which suggest that the composite likelihood containing a single sub-likelihood with
$\Ncl^\ast=1$ gives the best modelling. The $p$-value of the FS-CL test at $\Ncl^\ast=1$ equals 0.04, while the
$p$-values for the Wald, LSSB and LSSBw tests are 0.08, 0.09, and 0.03 respectively. At significance level 0.05,
the FS-CL and LSSBw tests suggest the null hypothesis be correctly rejected, while the other tests cannot reach
the same conclusion.

\begin{table}[htb]
\centering
\scalebox{1}{
\begin{tabular}{ cccccccccccccc}\hline
& \multicolumn{4}{c}{FS-CL}  & Wald  & LSSB & LSSBw \\
$\Ncl^\ast$    & 1    & 2  & 3 & 4 &   5    &      &     &             \\
\hline
CL-BIC          &  676 & 739 &  772 & 801 &   &  & \\
$p$-value    &0.04  &0.02&  0.06&  0.08 & 0.08 & 0.09 & 0.03 \\ \hline
\end{tabular}}
\caption{CL-BIC values with respect to $\Ncl^\ast$ ranking from 1 to 4,
as well as the $p$-values of the FS-CL test and the Wald type tests describe in Section~\ref{sec:tests}, for the 5
dependent SNPs. The CL-BIC values are computed using (\ref{TIC}), and the $p$-values of the FS-CL test are
acquired from the permutation null distribution of the test statistics as described in Section~\ref{sec:nulldistribution}. }
\label{Table5}
\end{table}

\section{Conclusion and discussion}
\label{sec:conclusion}

Building on the well-established composite likelihood estimation framework, we have developed a method of
simultaneous composition rule selection and group difference testing in multivariate parametric models for
high-dimensional data. The method is particularly useful for multiple genotype-phenotype association testing in
genome-wide association study.
It constructs sparse composite likelihood by including a small number of informatively selected
sub-likelihoods, while dropping redundant or noisy sub-likelihoods that do not contribute to explaining
the group difference or genomic association. The procedure is implemented by our forward search and test algorithm
which progressively includes useful sub-likelihoods by step-up maximizations of the bootstrap estimated
power. In all our numerical experiments, the resultant FS-CL test has higher power than
the composite likelihood based Wald, LSSB and LSSBw tests, with remarkable power gains when the model
complexity increases.

The FS-CL method has been applied to analyze a case-control dataset for GWAS, obtained from Australian Breast Cancer
Family Study, under the multivariate latent Gaussian framework studied by \cite{han2012composite}.
The FS-CL test enables us to conclude about the significant overall association between particular SNPs
and breast cancer, while the other Wald-type tests often cannot identify any such association.
Based on the performance of the FS-CL test in our numerical experiments, we believe the FS-CL procedure can be
a valuable tool for simultaneous model selection and group difference (or genomic association) testing.

Generalizing the FS-CL procedure is possible, which may lead to further improvements in terms of estimation accuracy
and test power. First, recall that the composite likelihood function (\ref{comp_lik}) admits only binary weights with
$w \in \{0,1\}^{\Ncl}$. A natural implication of this framework is the sparsity of the resulting
likelihood composition (and the induced parameter space). Developing a continuous weighting scheme
for strengthening informativeness of the selected sub-likelihoods may further decrease the MCLE variance and
increase the test power.
So far the overall model complexity in our framework is kept under control by running a forward step-up
procedure for including informative sub-likelihoods progressively, and by limiting the maximum
number of sub-likelihoods $\Ncl^\ast$ (cf. Section~\ref{sec:modelcomplexity}).
In using continuous and sparse weights, however, the model complexity control may be better achieved by a
sparsity-inducing smoothness penalization scheme for the weights, in the same spirit of the well established
high-dimensional variable selection procedures in the regression literature (see e.g. \cite{buhlmann2011statistics}).

\subsection*{Appendix: Density for the sum of ordered gamma variables}
Let $Y_1,\cdots, Y_K$ be $K$ i.i.d. $\Gamma(2^{-1}p^\prime,1)$ random variables. Define
$S_k=\sum_{j=1}^k Y_{(j)}$, $k=1,\dots, K$, with $Y_{(1)}\geq \cdots\geq Y_{(K)}$ being the reverse
order statistics of $Y_1, \cdots, Y_K$. Let $f_S(t|k)$ be the density of $S_k$. It is easy to show that the asymptotic
density of the FS-CL statistic $T_{\fscl}$ following (\ref{disktppclfs}) is $f_T(t|\Ncl^\ast)=
2^{-1}f_S(2^{-1}t|k=\Ncl^\ast)$.
Let $g_{\nu}(x)$, $G_{\nu}(x)$ denote the density function and distribution function
of a $\Gamma(\nu,1)$ distribution, respectively.
\cite{alam1979distribution} derived an analytic form for the density of $S_k$ which is given as
\begin{align}
f_S(t|k)= \frac{k {K\choose k}}{\Gamma(2^{-1}p^\prime+1)^{K-k}}\sum_{r=0}^\infty
(-1)^r d_{r,K-k} l_{r,k} \frac{1}{r!} g_{2^{-1}p^\prime K+r}(t),
\end{align}
where
$l_{r,k}=\int_0^\infty
x^{2^{-1}p^\prime (K-k)+r}(1-G_{2^{-1}p^\prime}(x))^{k-1}g_{2^{-1}p^\prime}(x)\text{d}x$;
$d_{r,s}$ is computed recursively as
\begin{align*}
d_{0,s}=1, \quad d_{r,1} =\frac{p^\prime}{p^\prime+2r}, \quad
d_{r+1,s}=\frac{p^\prime s}{2}\sum_{j=0}^r {r \choose j}
\left(2^{-1}p^\prime+1+r-j \right)^{-1} d_{j,s-1}, \ \  s>1.
\end{align*}
Thus, the asymptotic density of the test statistic $T_{fscl}$  following (\ref{disktppclfs}) is given by
\begin{align} \label{null}
f_T(t|k)= \frac{k {K\choose k}}{2\Gamma(2^{-1}p^\prime+1)^{K-k}}\sum_{r=0}^\infty
(-1)^r d_{r,K-k} l_{r,k} \frac{1}{r!} g_{2^{-1}p^\prime K+r}(2^{-1}t),
\end{align}
which is a mixture of $\chi^2$ distributions with varying degrees of freedom.

\bibliographystyle{abbrvnat} 

\bibliography{Bibliography} 

\end{document}